\documentclass[aps,prb,twocolumn]{revtex4}
\usepackage{amsmath}    
\usepackage{amsopn}
\usepackage{graphicx}   
\usepackage{verbatim}   
\usepackage{color}      
\usepackage{subfigure}  
\DeclareMathOperator{\Tr}{Tr}
\begin{document}
\title{Formation of magnetic skyrmions with tunable properties in PdFe bilayer deposited on Ir(111)}
\author{E. Simon$^{1}$, K. Palot\'as$^{1}$, L. R\'ozsa$^{1}$, L. Udvardi$^{1,2}$ and L. Szunyogh$^{1,2}$}
\affiliation{$^1$Department of Theoretical Physics, Budapest University of Technology and Economics, Budafoki \'{u}t 8., H-1111 Budapest, Hungary\\
$^2$Condensed Matter Research Group of Hungarian Academy of Sciences, Budapest University of Technology and Economics, Budafoki \'ut 8, H-1111 Budapest, Hungary}

\date{\today}

\begin{abstract}
We perform an extensive study of the spin-configurations in a PdFe bilayer on Ir(111) in terms of ab initio and spin-model calculations. 
We use the spin-cluster expansion technique to obtain spin model parameters, and solve the Landau-Lifshitz-Gilbert equations at zero temperature. 
In particular, we focus on effects of layer relaxations and the evolution of the magnetic ground state in external magnetic field. 
In the absence of magnetic field, we find a spin-spiral ground state, while applying external magnetic field skyrmions are generated in the system. Based on energy calculations of frozen spin configurations with varying magnetic field we obtain excellent agreement for the phase boundaries with available experiments. We find that the wave length of spin-spirals and the diameter of skyrmions decrease with increasing inward Fe layer relaxation which is correlated with the increasing ratio of the nearest neighbor Dzyaloshinskii-Moriya interaction and the isotropic exchange coupling, $D/J$. Our results also indicate that the applied field needed to stabilize the skyrmion lattice increases when the diameter of individual skyrmions decreases. Based on our observations, we suggest that the formation of the skyrmion lattice can be tuned by small structural modification of the thin film.
\end{abstract}

\maketitle

\section{Introduction}

The concept of skyrmions (Sk) was originally introduced in nonlinear field theory \cite{skyrme} and then generally used   as quasi-particle excitations in different fields of physics and mathematics. Skyrmions were first measured in quantum-Hall ferromagnets and the crystallization of skyrmions was also realized. \cite{hall1,hall2}
Magnetic skyrmions are chiral spin structures that are topologically protected, therefore, they are relatively stable against thermal fluctuations. These magnetic spin structures were first observed experimentally in the bulk MnSi \cite{muhlbauer,pappas} and examined theoretically by using a mean-field model. \cite{butenko} At low temperature and low magnetic field, MnSi develops a helical magnetic structure. Due to the large ferromagnetic exchange interaction a uniform spin alignment would, in principle, be favored, however
the lack of inversion symmetry in cubic B20 MnSi \cite{muhlbauer} results in 
Dzyaloshinskii-Moriya (DM) interactions \cite{Dzyaloshinsky,Moriya} that induce the helical structure. 
By applying an external magnetic field, a columnar skyrmion lattice (SkX) develops which is referred to as the A phase of bulk MnSi.  
Similar skyrmion lattice formation was observed in a thin film of B20-type Fe$_{0.5}$Co$_{0.5}$Si, where the thickness of the film was less than the wave length of the helical spin structure. \cite{yu} As clear from the phase diagrams \cite{muhlbauer,yu} and other theoretical studies \cite{butenko},  external magnetic field and finite temperature are needed to stabilize the skyrmion lattice, however, the ranges of the stabilizing field and temperature are relatively narrow. It should be noted that in contrast to the previous results, for epitaxial FeGe(111) films the skyrmion phase has been stabilized up to 250 K. \cite{huang}

Application of magnetic skyrmions in ultrathin films in spintronic devices is an appealing issue,~\cite{fert} thus, the conditions of formation and the properties of the magnetic skyrmions are widely studied. 

 In a recent experimental work the formation of individual skyrmions has been observed in a PdFe bilayer deposited on Ir(111) surface. \cite{romming} A spin-spiral ground state has been revealed at temperature of 8 K, while  applying a relatively small external magnetic field (B=1 T) skyrmions were created. The diameter of the skyrmions, $\sim 5-6$ nm, was much larger than that of the nanoskyrmions in a single Fe atomic layer on Ir(111), $\sim 1$ nm. \cite{heinze} However, this nanoskyrmion lattice most possibly manifests a phase different from that observed in PdFe/Ir(111), because it is stable even without an external magnetic field. 
 Skyrmion magnetic structures including a $B-T$ phase diagram were first reported in terms of a combined ab initio and spin-model study for the ordered FePt monolayer deposited on Pt(111). \cite{polesya}
 The stability, the structural and dynamic properties of skyrmions in ultrathin films were very recently investigated theoretically in several contexts by Dup\'e {\em et al.}  \cite{dupe}

In the present work we study the magnetic properties of the PdFe bilayer on Ir(111) surface using first principles calculations. We use a spin-cluster expansion (SCE) technique combined with the relativistic disordered local moment (RDLM) scheme to obtain parameters of a spin model. Using these parameters, the magnetic ground state with and without external magnetic field is examined by spin-dynamic simulations. We highlight the role of the layer relaxations on the obtained magnetic interactions within the Fe layer. We find that the geometry of the magnetic interface considerably affects both the isotropic exchange and the DM interactions, thus the formation of the skyrmion state. With increasing inward layer relaxation, the ratio of DM and isotropic exchange is increased. In agreement with Fert et al. \cite{fert}, we report reduced skyrmion sizes with a large ratio of the DM interaction and isotropic exchange. We also find that the applied field needed to stabilize the skyrmion lattice increases with decreasing diameter of the skyrmions. This means that the formation of the skyrmions is tunable 
by inducing small structural changes in the system.

\section{Computational details}

Based on the adiabatic decoupling of fast electronic fluctuations from the slow transversal motion of spins, \cite{antropov} the spin-system can be described in terms of a classical spin-model. We use a generalized Heisenberg model,
\begin{equation}
\mathcal{H}=-\frac{1}{2}\sum_{i \ne j}\vec{s}_{i}\mathbf{J}_{ij}\vec{s}_{j}+\sum_{i}\vec{s}_{i}\mathbf{K}_{i}\vec{s}_{i} - \sum_i m_i \vec{s}_{i} \vec{B}_{ext},
\label{genHeis}
\end{equation} 
where $\vec{s}_{i}$ represents the direction of the magnetic moment at site $i$, $\vec{m}_{i} = m_i \vec{s}_{i} $. The first term of Eq.~(\ref{genHeis}) stands for the exchange contribution with tensorial exchange coupling, $ \mathbf{J}_{ij}$,\cite{udvardiRTM} which can be decomposed into an isotropic component, $J_{ij}\mathbf{I}$ with $J_{ij}=\frac{1}{3}\Tr\mathbf{J}_{ij}$, an antisymmetric component $\mathbf{J}_{ij}^{A}=\frac{1}{2}(\mathbf{J}_{ij}-\mathbf{J}_{ij}^{T})$, and a traceless symmetric part $\mathbf{J}_{ij}^{S}=\frac{1}{2}(\mathbf{J}_{ij}+\mathbf{J}_{ij}^{T})-J_{ij}\mathbf{I}$, where the superscript $T$ denotes the transpose of a matrix and $\mathbf{I}$ is the unit matrix. The first, isotropic component describes the Heisenberg interaction. The energy term -$\vec{s}_{i}\mathbf{J}_{ij}^{A}\vec{s}_{j}=\vec{D}_{ij}(\vec{s}_{i}\times \vec{s}_{j})$ corresponds to the DM interaction with $\vec{D}_{ij}$ being the DM vector \cite{Dzyaloshinsky,Moriya}. The symmetric part of the exchange term is the two-site magnetic anisotropy and the second term of Eq.~(\ref{genHeis}) comprises the on-site anisotropy with the  anisotropy matrix $\mathbf{K}_{i}$. 
The third term of Eq.~(\ref{genHeis}) is the Zeeman energy of the spin-moments of magnitude $m_i$ in the presence of an external field, $\vec{B}_{ext}$.
Neglecting self-consistent longitudinal spin-fluctuations, two methods are used in the literature to map the energy from first-principles calculations to the generalized spin Hamiltonian in Eq.(\ref{genHeis}). The relativistic torque method \cite{udvardiRTM,ebertRTM} makes use of infinitesimal rotations around specific magnetic configurations, mostly around the ferromagnetic state (FM). For this reason, if the ground state is far from the FM state, the interaction parameters might be inconsistent with the original ground state. The spin-cluster expansion (SCE) technique developed by Drautz and F\"ahnle \cite{DrautzSCE1,DrautzSCE2} provides a systematic parametrization of the adiabatic energy of the spin-system. The SCE method was combined with the relativistic disordered local moment scheme (RDLM) \cite{gyorffy,stauntonRDLM1, stauntonRDLM2, buruzsRDLM}, and this combination  gives a proper tool to determine the parameters of the spin Hamiltonian in Eq.(\ref{genHeis}) from the paramagnetic state.\cite{szunyoghSCE,adeak,simon-fe5d}
Note that by using the SCE method  no \textit{a priori} information about the magnetic ground state is needed.

In terms of the screened Korringa-Kohn-Rostoker (SKKR) method \cite{szunyoghKKR,zellerKKR} we performed self-consistent calculations of a PdFe bilayer deposited on Ir(111) surface.  We employed the scalar-relativistic DLM approach \cite{gyorffy} to obtain the electronic structure in the paramagnetic state. The local spin-density approximation (LSDA) as parametrized by Vosko et al. \cite{vosko} was used within the atomic sphere approximation with an angular momentum cut-off of $l_{max}=3$. The energy integrals were performed by sampling $16$ points on a semicircle contour in the upper complex semi-plane.
To model the geometry of the system, the in-plane lattice constant of the Ir, $a_{2D}$=2.715\AA\;was chosen, and fcc growth was assumed 
for both the Fe and the Pd layers. It was indeed confirmed by Dup\'e {\em et al.}  \cite{dupe} that the fcc growth is lower in energy as compared to the hcp growth. We performed geometry optimization in terms of  VASP calculations \cite{kresseVASP1,kresseVASP2,hafnerVASP} by modeling the PdFe/Ir(111) system as a slab of nine layers (Pd + Fe +7 layers Ir). This resulted in a relaxation of $-5$\% of the Fe layer. In order to study the effect of structural modifications, in our SKKR calculations we considered inward relaxations of the Fe layer ranging from $-5$\% to $-10$\%. Following the self-consistent calculations, for each values of the Fe layer relaxation we derived the parameters of the spin-model Eq.~(\ref{genHeis}) in terms of the SCE-RDLM method.\cite{szunyoghSCE} Note that since in the self-consistent DLM state the local spin-polarization of the Pd and Ir atoms disappeared, no spin-model parameters were calculated for these atoms, therefore we considered only Fe spins in the spin-model Eq.~(\ref{genHeis}).   

For finding the ground state, we performed zero temperature (deterministic) Landau-Lifshitz-Gilbert spin-dynamics simulations which describes the motion of the localized magnetic moments, \cite{rozsa1, rozsa2}  
\begin{equation}
\frac{\partial \vec{m}_{i}}{\partial t}=-\frac{\gamma}{1+\alpha^{2}}\vec{m}_{i}\times \vec{B}_{i}-\frac{\alpha \gamma}{(1+\alpha^{2})m_{i}}\vec{m}_{i}\times (\vec{m}_{i}\times \vec{B}_{i}),
\end{equation}
where $\alpha$ is the Gilbert damping parameter, $\gamma=2\mu_{B}/\hbar$ is the gyromagnetic ratio and the effective field, $\vec{B}_{i}$, is obtained from the generalized Hamiltonian, Eq. (\ref{genHeis}), 
\begin{equation}
 \vec{B}_{i} = - \frac{1}{m_i} \frac{\partial \cal{H}}{\partial \vec{s}_i} =  
 \frac{1}{m_i} \sum_{j (\ne i)}  {\bf J}_{ij}  \vec{s}_j -
 \frac{2}{m_i}  {\bf K}_{i}  \vec{s}_i+ \vec{B}_{ext} \, .
\end{equation}
  We used a two--dimensional lattice of $128\times128$ sites populated by classical spins with periodic boundary condition and considered the full tensorial exchange interactions and the on-site anisotropy term when calculating the effective field. Each simulation was initialized at a random spin configuration and continued until the absolute difference in the energy of the spin system between two steps reached the value of $10^{-5}$ mRy. In the simulations $\alpha=0.01$ was used with a sufficiently small time step to ensure a stable search for the ground state.

\section{Results}

\begin{figure}[t!]
\begin{center}
\includegraphics[scale=0.6]{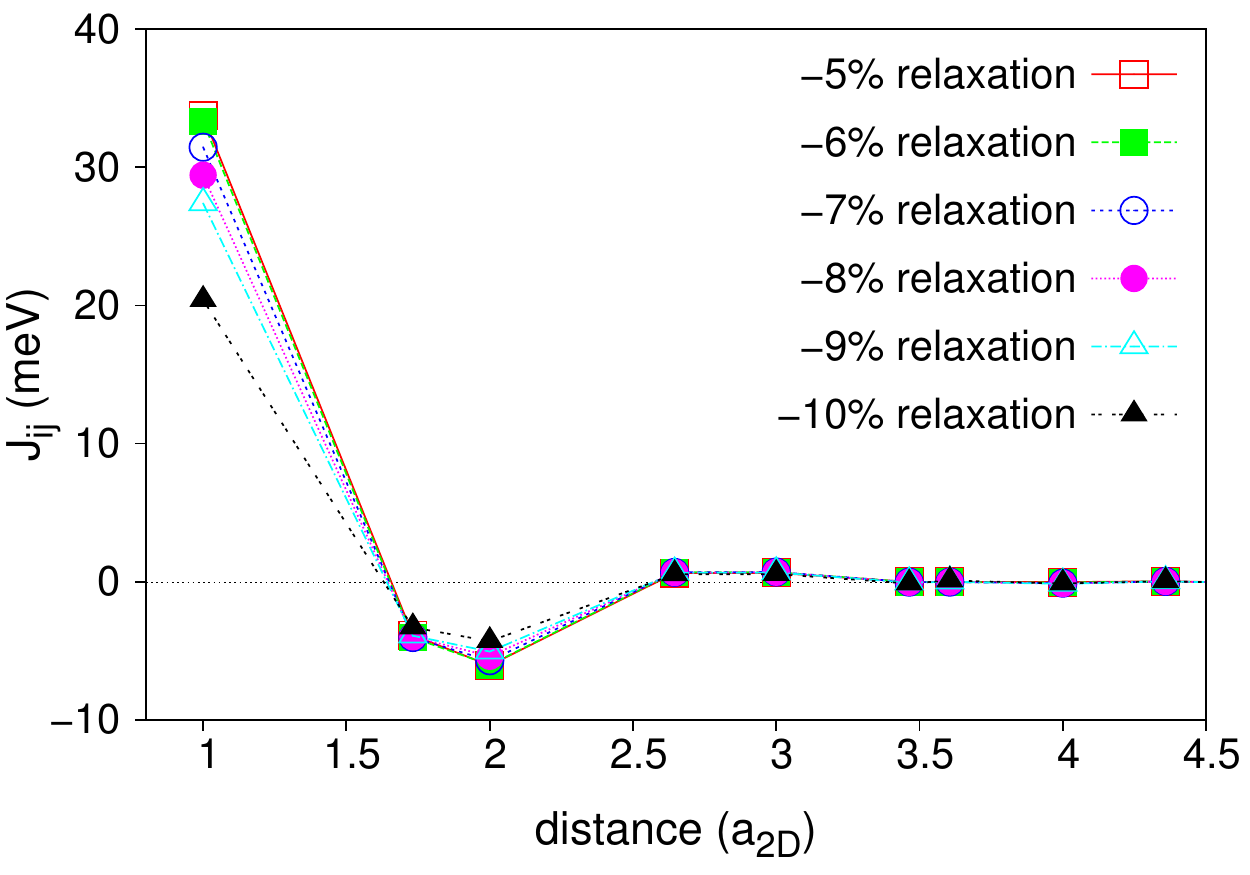}
\caption{(Color online) Calculated Fe-Fe isotropic exchange interactions for PdFe/Ir(111) as a function of the inter-atomic distance measured in units of the in-plane lattice constant ($a_{2D}$) for different Fe layer relaxations.}
\label{calc-jij}
\end{center}
\end{figure}

Fig. \ref{calc-jij} shows the calculated Fe-Fe isotropic exchange interaction as a function of the inter-atomic distance for all considered layer relaxations. According to Eq.~(\ref{genHeis}), the positive sign of the exchange interaction means ferromagnetic (FM) coupling and the negative sign refers to antiferromagnetic (AFM) coupling. For all considered layer relaxations the nearest neighbor exchange interactions are FM and they gradually decrease with increasing inward layer relaxation. In the second and third shells the exchange interactions are AFM that turn back to FM from the fourth shell. Interestingly, from the second shell the interactions just slightly depend on the inward layer relaxation.

\begin{figure}[t!]
\begin{center}
\includegraphics[scale=0.6]{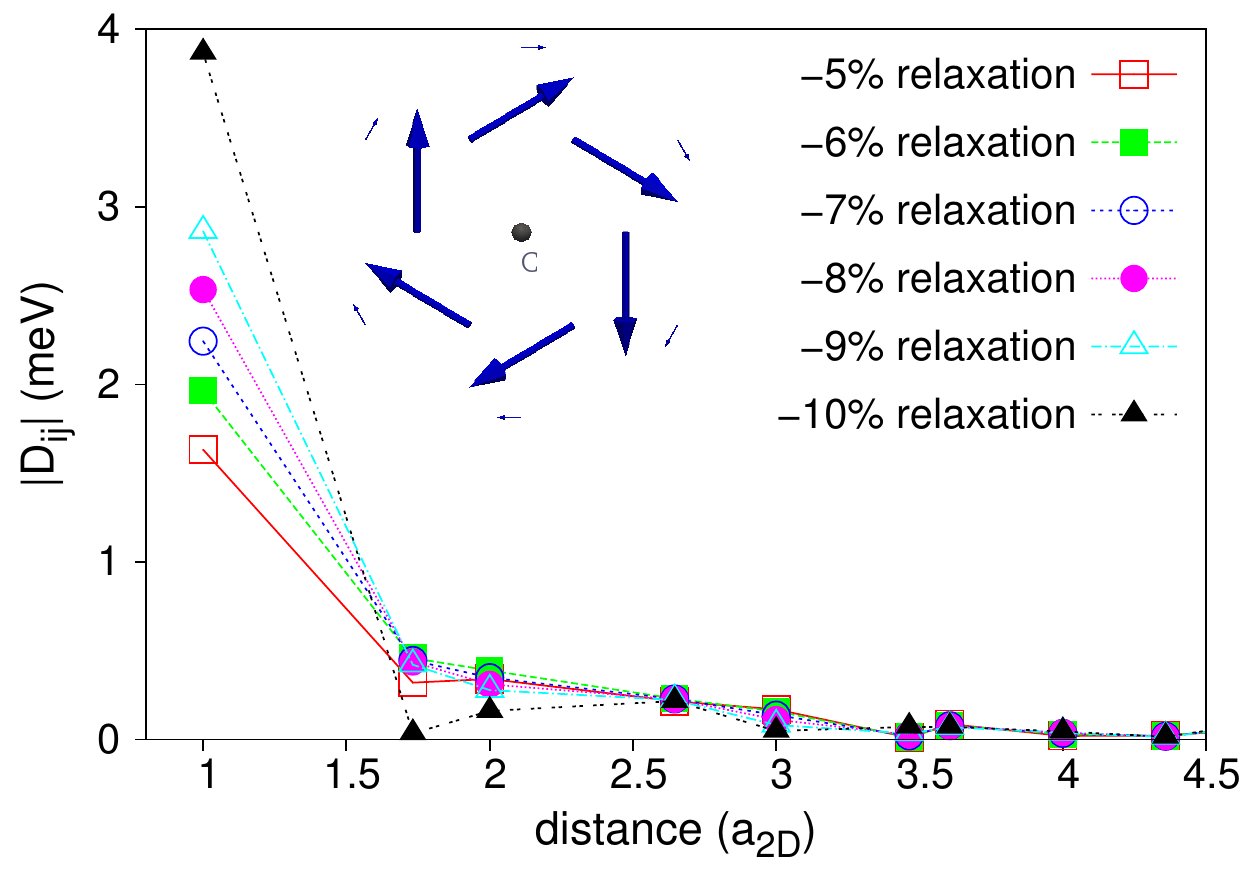}
\caption{(Color online) Magnitudes of the Fe-Fe DM vectors for PdFe/Ir(111) as a function of the inter-atomic distance measured in units of the in-plane lattice constant ($a_{2D}$) for different Fe layer relaxations. The inset shows a sketch of the in-plane components of the calculated DM vectors between a central Fe atom ($C$) and its nearest and next nearest Fe neighbors at $-5$\% relaxation.}
\label{dm-jij}
\end{center}
\end{figure}

As mentioned above, beside the isotropic exchange interaction the DM interactions can play an important role in the formation of complex magnetic ground states in ultrathin films. \cite{bodeSS,heinze,dupe,polesya,simon-fe5d} In Fig. \ref{dm-jij} the magnitudes of the DM vectors between the Fe atoms are shown as a function of the inter-atomic distance for all considered layer relaxations. It can be seen that the largest magnitude of the DM vector is found for the first Fe neighbors, for further shells the DM vectors are much smaller in magnitude. It is also clearly seen that the magnitude of the DM vectors for the first shell increases with increasing inward layer relaxation. Since the DM interaction prefers non-collinear alignment of the magnetic moments, the large DM vectors in the first shell indicate the formation of a spin-spiral structure in the Fe layer as the magnetic ground state.

It should be noted that due to the $1/2$ factor in the first term of the spin Hamiltonian in Eq.~(\ref{genHeis}) our spin model parameters are twice as large as in Ref.\ \onlinecite{dupe}. Taking this into account, at $-5$\% relaxation we obtained $J_1=16.87$~meV and  $D_1=0.82$~meV for the nearest neighbor isotropic exchange interaction and magnitude of the DM vectors, respectively, while in Ref.\ \onlinecite{dupe} $J_1=14.7$~meV and $D_1= 1.0$~meV were reported. Considering the quite different theoretical approaches, this means a very good agreement between the two calculations.

 The inset of Fig. \ref{dm-jij} shows the in-plane projection of the  DM vectors for the nearest and second nearest neighbors in case of $5\%$ inward layer relaxation. Obviously, the orientations of the DM vectors in a given shell are consistent with the $C_{3v}$ point group symmetry of the system. Note that the DM vectors transform as axial vectors. \cite{Dzyaloshinsky,Moriya} Our calculations evidence that the in-plane components of the DM vectors  in the first and second shells are much larger than the out-of plane components implying an out-of plane rotation of the spins. For the first shell, the magnitude of the in-plane component is $D_{\parallel}=1.58$ meV, whilst the out-of plane component is $D_{\perp}=0.41$ meV. Similar behavior of the DM vector components can be obtained for all considered layer relaxations.
Note that the Fe-Fe isotropic exchange and DM interactions in the first three shells are increased in magnitude due to the presence of the Pd overlayer on the Fe/Ir(111). For the nearest neighbor interactions this increase was about 30\%. A similar effect was found by Dup\'e et al. \cite{dupe}

First, we estimated the magnetic ground state of the system by calculating the Fourier transform of the exchange matrices, $\mathbf{J}(\vec{q})$. For a spin system described by the first term in the Heisenberg Hamiltonian Eq. (\ref{genHeis}), the energy of a spin-spiral with a wave-vector $\vec{q}$ is given by the minimum eigenvalue of $-\mathbf{J}(\vec{q})$ or equivalently, by the maximum eigenvalue of $\mathbf{J}(\vec{q})$. \cite{simon-fe5d,ondracek,adeak,kudrnovsky}  A uniaxial magnetic anisotropy, $K \cos^2 \theta_i$ with $\theta_i$ being the polar angle of the magnetization at site $i$,  adds $K/2$ per unit cell to the energy of such a spin-spiral, while -$\mid \!\! K \!\!\mid$ to that of the FM state of minimal energy. Note that for $-5$\% relaxation of the Fe layer we calculated $K = -0.5$~meV being considerably  smaller than the leading isotropic exchange and DM interactions. 
Neglecting on-site anisotropy, a maximum of the eigenvalues of $\mathbf{J}(\vec{q})$ at the center of the Brillouin zone, i.e. at the $\overline{\Gamma}$ point thus means a ferromagnetic ground state, whilst a maximum located at a general $\vec{q}$ vector of the Brillouin zone corresponds to a more complex magnetic ground state. For all considered Fe layer relaxations, the maximum was found close to the $\overline{\Gamma}$ anticipating a spin-spiral ground state of large wave length.

From the spin-dynamics simulations we obtained a spin-spiral ground state in accordance with the estimation based on  $\mathbf{J}(\vec{q})$. The estimated wave length from the maximum eigenvalue of $\mathbf{J}(\vec{q})$ and the wave length obtained from the spin-dynamics simulations are in remarkably good agreement with each other, similarly as found in the Fe/Os(0001) system.\cite{simon-fe5d} As can be seen in Table \ref{tab:wave}, the wave length of the spin-spiral ($\lambda$) decreases with increasing inward layer relaxation. This can be correlated with the ratio of the magnitude of the nearest neighbor DM vectors and the isotropic exchange interaction ($D/J$) also presented in Table \ref{tab:wave}. There is almost an inverse proportionality between $\lambda$ and $D/J$ as can be obtained from a simple analytic estimation in the small wave number limit.\cite{bodeSS} It should be noted that for $-5$\% relaxation, which is the energetically favored geometry from the VASP method, the calculated wave length of 6.8 nm is in excellent agreement with the experimentally measured spin-spiral period of about 6 to 7 nm. \cite{romming}

\begin{table}[ht!]
\begin{center}
\begin{tabular}{c c c c}
\hline
 Relaxation  & & $D/J$  &  $\lambda$ (nm)\\
\hline
\hline
$-5$\% &  &  0.05   &   6.8    \\
$-6$\% &  & 0.06   &   5.4    \\
$-7$\% &  &  0.07   &   4.7    \\
$-8$\% &  &  0.09   &   4.1    \\
$-9$\% &  & 0.10   &   3.6    \\
$-10$\% & &   0.19   &   2.4
\end{tabular}
\caption{Ratio of the magnitudes of the nearest neighbor Fe-Fe DM vector ($D$) and isotropic exchange coupling ($J$) as well as the wave length ($\lambda$) of the ground state spin-spiral for each value of Fe layer relaxation in PdFe/Ir(111).}
\label{tab:wave}
\end{center}
\end{table}

\begin{figure}[ht!]
\begin{center}
\includegraphics[scale=0.6]{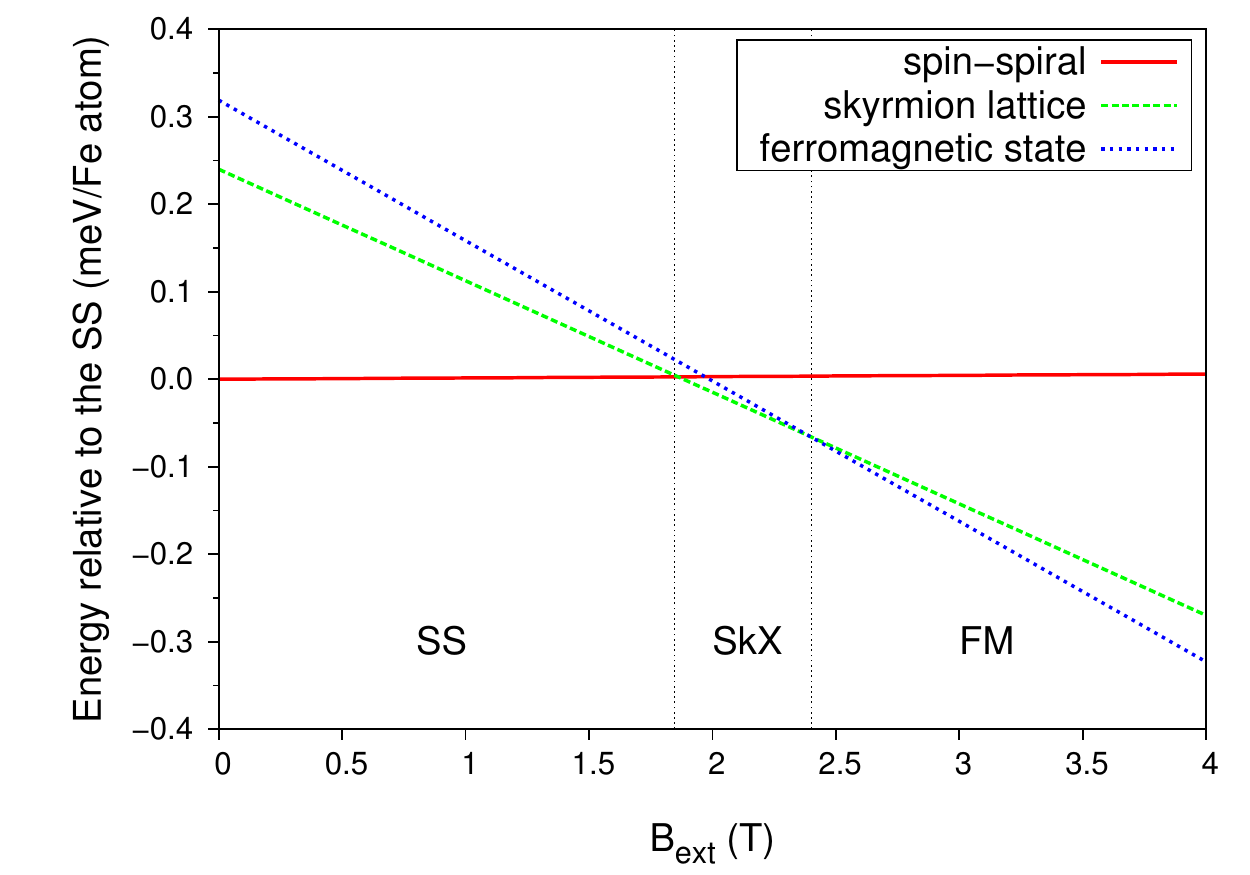}
\caption{(Color online) Energies of frozen spin-spiral (SS), selected skyrmion phase (SkX), and ferromagnetic (FM) spin structures in external magnetic field ($B_{ext}$) perpendicular to the surface for $-5$\% relaxation of the Fe layer in PdFe/Ir(111). All the energies correspond to the 128$\times$128 lattice used for the spin-dynamics simulations. The vertical lines denote the phase boundaries between the three different states.}
\label{energy}
\end{center}
\end{figure}

\begin{figure}[t!]
\begin{center}
\includegraphics[scale=0.6]{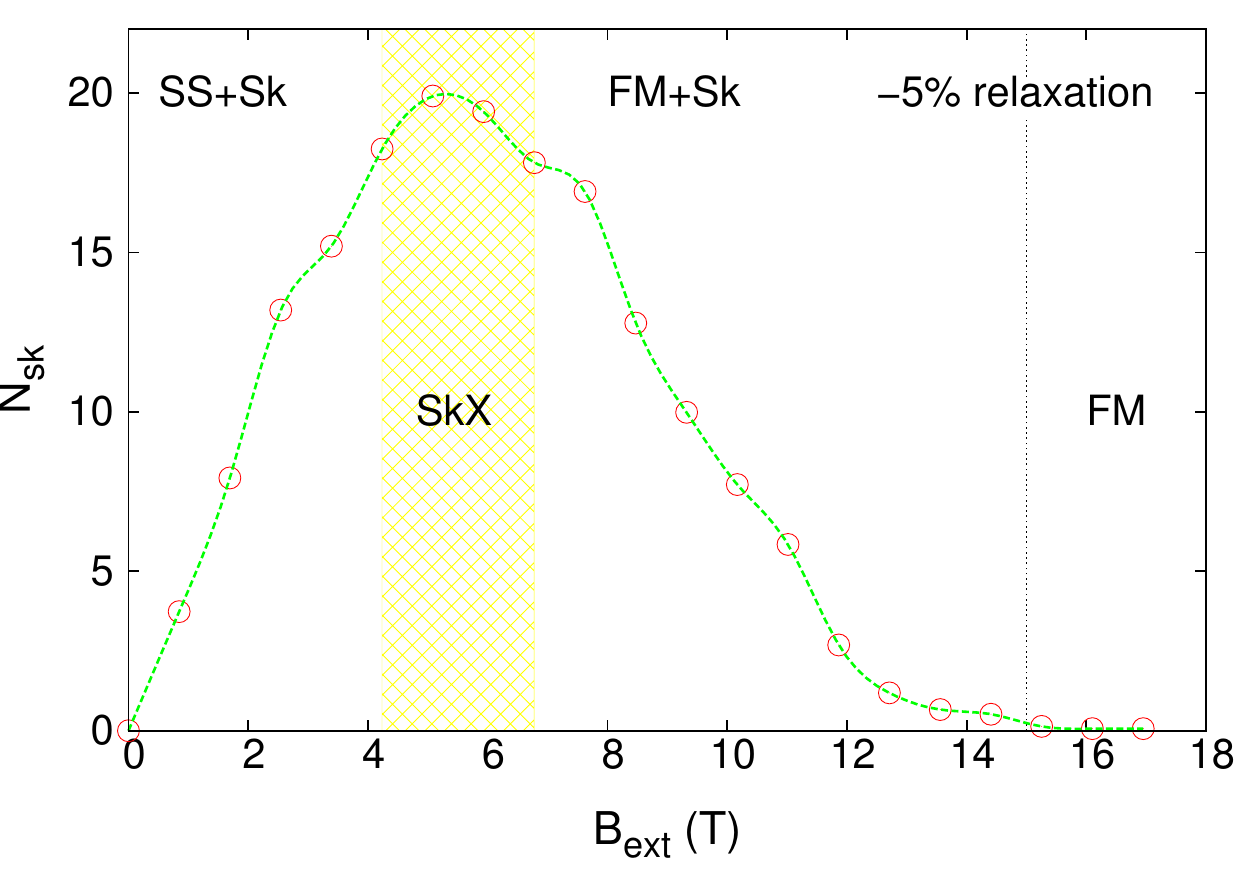}
\includegraphics[scale=0.6]{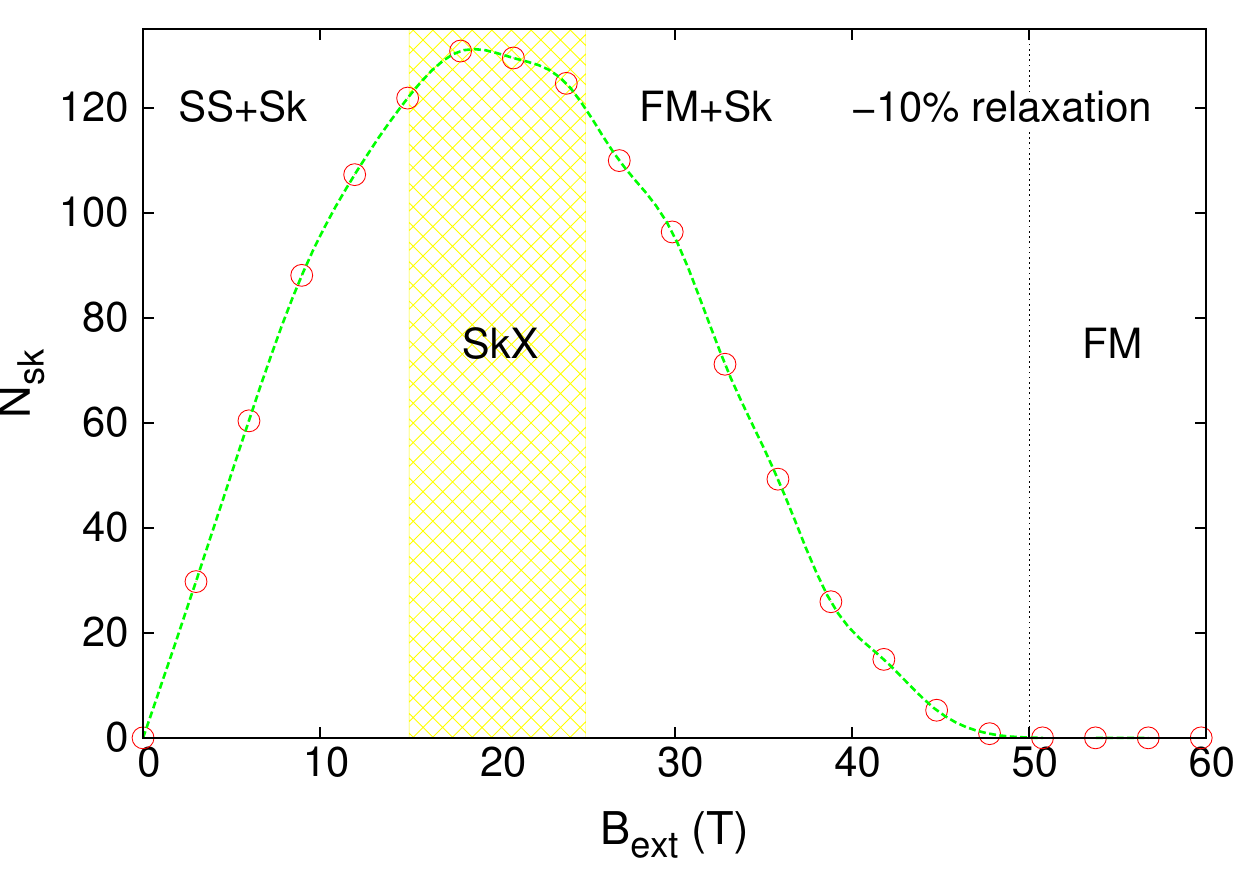}
\caption{(Color online) Calculated skyrmion numbers ($N_{sk}$) on a $128\times128$ lattice at zero temperature as a function of external magnetic field ($B_{ext}$) in case of $-5$\% and $-10$\% relaxations of the Fe layer in PdFe/Ir(111). The points represent skyrmion numbers averaged over 15 independent spin-dynamic simulations and the lines denote an interpolated curve between the points. The magnetic phases as described in the text are indicated.} 
\label{sknum}
\end{center}
\end{figure}

When an external magnetic field, $B_{ext}$, is applied perpendicular to the surface, at low temperature the spin-spiral structure can change to a 2D skyrmion lattice. In an external magnetic field the energy of the spin-spiral (SS), the skyrmion lattice phase (SkX) and the FM state is changing differently due to the different out-of-plane spin-component of these spin structures entering the Zeeman term of the energy. Assuming frozen magnetic configurations for the energetically favored geometry ($-5$\% relaxation), the energy dependence of the mentioned spin structures on the external magnetic field at zero temperature is shown in Fig. \ref{energy}.
In particular, for the skyrmion phase we considered the spin-structure with the maximum skyrmion number (see later in context to Fig. \ref{sknum}). 

At $B_{ext}=0$ the energy of the spin-spiral state is preferred and the highest energy is obtained for the ferromagnetic state. Since the net magnetization is zero in the spin-spiral state, its Zeeman energy is also zero, therefore, the spin-spiral energy is constant against $B_{ext}$. If the spins of the ferromagnetic state are parallel to the external field, the Zeeman contribution reduces the total energy by increasing  $B_{ext}$. The slope of the curve corresponds to $N m$, where $N$ is the number of lattice sites and $m$ is the size of the Fe moments. Similar decreasing energy can be observed in case of the skyrmion lattice. Here, the slope of the energy curve is smaller than in the FM state due to the smaller net out-of-plane component of the spins. With increasing external magnetic field the energy minimum changes first from spin-spiral to skyrmion lattice. Further increasing $B_{ext}$ leads to the saturation of the Fe moments in the ferromagnetic state.
The slope of the energy curve of the skyrmion lattice naturally depends on the actual skyrmion state, more precisely on the number and size of individual skyrmions in the system. Therefore, we calculated the energy curve of several different frozen skyrmion states as obtained from spin-dynamics simulations starting with different initial configurations and found that the SS-SkX intersection is in the range between 0.7 and 1.8 T and the SkX-FM intersection is between 2.4 and 3.2 T. These ranges of magnetic field are in good agreement with experimental observations.\cite{romming}

Fig. \ref{energy} should be regarded as an illustrative model  which gives a good qualitative picture about the origin of the magnetic phase transitions in PdFe/Ir(1111) at zero temperature. The phase boundaries can be determined more precisely by  performing spin-dynamics simulations with increasing $B_{ext}$. Here the spin-configurations are not frozen any more, but they are evolved to get the state with minimum energy for each value of  $B_{ext}$. By starting the simulations from different initial states, a statistics for the phase transitions can be obtained in terms of $B_{ext}$. From the spin-dynamics simulations we found the same range for the SkX-FM intersection as from the energy curves of frozen skyrmions. This finding suggests that the skyrmion lattice above about 3 T is metastable. It should be noted that Dup\'e et al. \cite{dupe} found magnetic phase transitions at much larger external fields as compared to the experiments.\cite{romming} 

According to our previous study, \cite{simon-fe5d} the spatial range of the magnetic interactions plays a crucial role in the formation of magnetic patterns in ultrathin films. In our spin-dynamics simulations we used 15 shells including isotropic exchange couplings, DM interactions and two-site anisotropies. We found that at least 4 shells are needed to obtain a spin-spiral configuration as the magnetic ground state and skyrmion formation under applying external magnetic field. This does not contradict to the observed strong relationship between the SS wave length and the ratio of nearest neighbor parameters, $D/J$, as it just highlights that further DM interactions are needed to decrease the energy of the spin-spiral below that of the FM state. Without DM interactions the magnetic ground state was ferromagnetic, in contrast to Dup\'e et al. \cite{dupe} who obtained SS ground state by neglecting spin-orbit interaction in their calculations.

The skyrmion phase is characterized by the skyrmion number (topological charge) defined by
\begin{equation}
N_{sk}=\frac{1}{4\pi}\int \vec{s}\cdot \left(\frac{\partial \vec{s}}{\partial x}\times \frac{\partial \vec{s}}{\partial y} \right) dx dy,
\end{equation}
where $\vec{s}$ is the direction of the local magnetization.  \cite{polesya} In case of topologically trivial magnetic structures, such as ferromagnetic, anti-ferromagnetic or spin-spiral states, the topological charge is zero. A single skyrmion holds the topological charge of $N_{sk}=1$, whilst for anti-skyrmion the charge is $N_{sk}=-1$.
In the PdFe/Ir(111) system, due to the clockwise rotation of the nearest neighbor Fe-Fe DM vectors, see inset of Fig.\ \ref{dm-jij}, skyrmions with winding number of one are formed. $N_{sk}$ then gives the number of skyrmions in the surface cell for which the integration is carried out. 

Based on the spin-configurations obtained from the spin-dynamics simulations at zero temperature we determined the number of skyrmions as described in Ref.~\onlinecite{polesya}. We found that averaging over 15 independent spin-dynamics simulations being started from random spin configurations is sufficient to stabilize the value of $N_{sk}$ for any layer relaxation and external magnetic field. Note that only about 500 time steps were sufficient to reach convergence of $N_{sk}$ as opposed to at least $10^4$ time steps for a precise determination of the ground state. Fig. \ref{sknum} shows the skyrmion number as a function of the external magnetic field for two different layer relaxations of $-5$\% and $-10$\%. As can be seen, without $B_{ext}$ the skyrmion number is zero, corresponding to the spin-spiral ground state. 

According to the variation of $N_{sk}$ against $B_{ext}$, we attempt to identify four different magnetic phases of the system. When increasing $B_{ext}$ from zero to a certain value, the skyrmion number is gradually increasing. In this phase individual skyrmions coexist with spin-spirals, therefore, we call it as a mixed spin-spiral and skyrmion, SS+Sk, state. With larger external magnetic field the skyrmion number saturates and there is a range of  $B_{ext}$ where  $N_{sk}$ is just slightly changed. Depending on the actual shape of the $N_{sk}(B_{ext})$ curves, this phase corresponds to the skyrmion lattice SkX, and it is defined as $N_{sk}>(0.9-0.95) \times N_{sk}^{max}$, where $N^{max}_{sk}$ is the maximal number of skyrmions. At even larger $B_{ext}$ the skyrmion number is decreasing because the skyrmions are saturated to the ferromagnetic state: this phase is a mixed ferromagnetic and skyrmion state, FM+Sk. Finally at a sufficiently large magnetic field the skyrmion number vanishes again when the ferromagnetic phase is reached.

\begin{figure*}[t!]
\centering
\subfigure[\label{fig:r5-nobext}] {\includegraphics[scale=0.14]{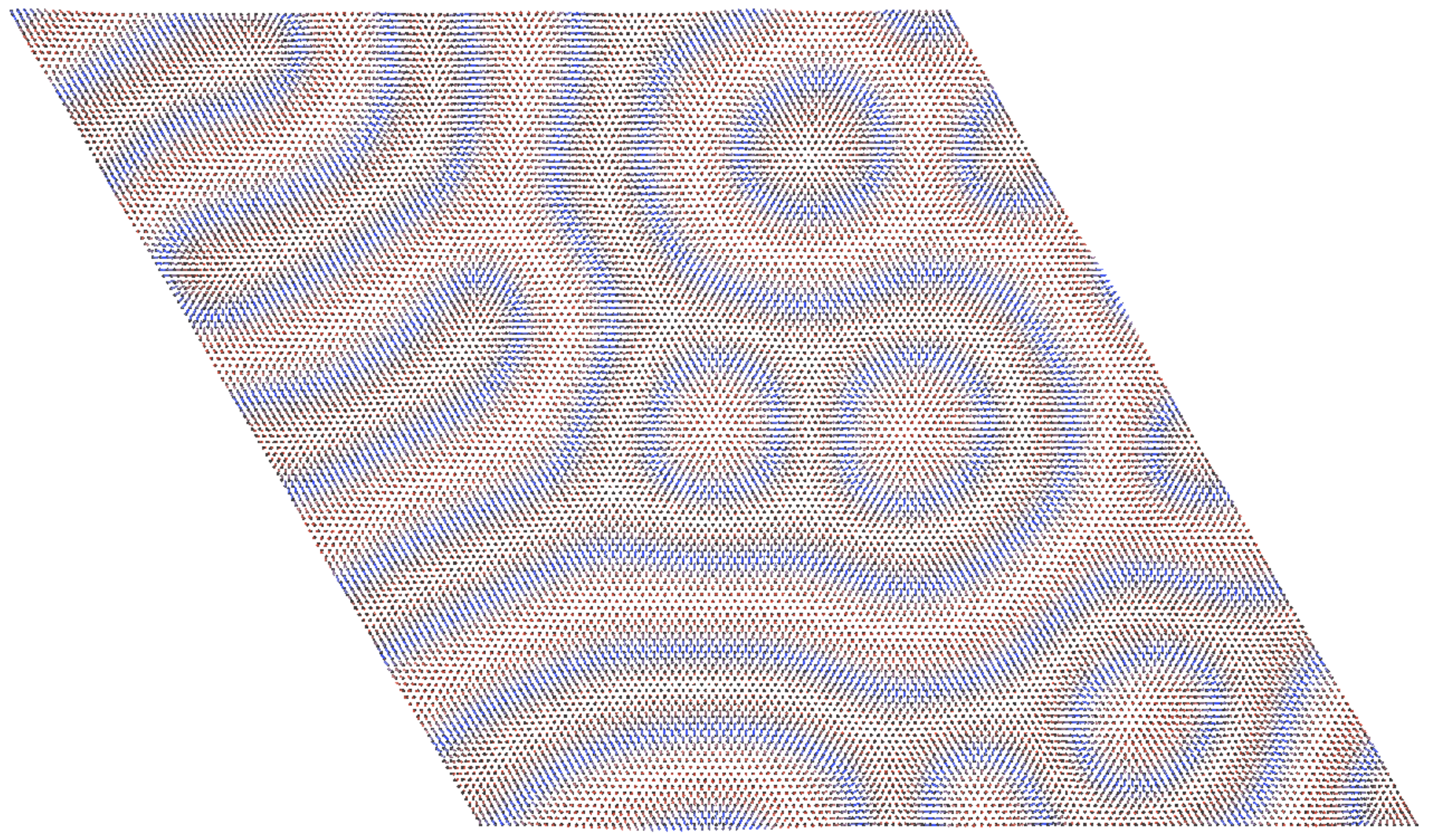}}
\subfigure[\label{fig:r5-bext}] {\includegraphics[scale=0.14]{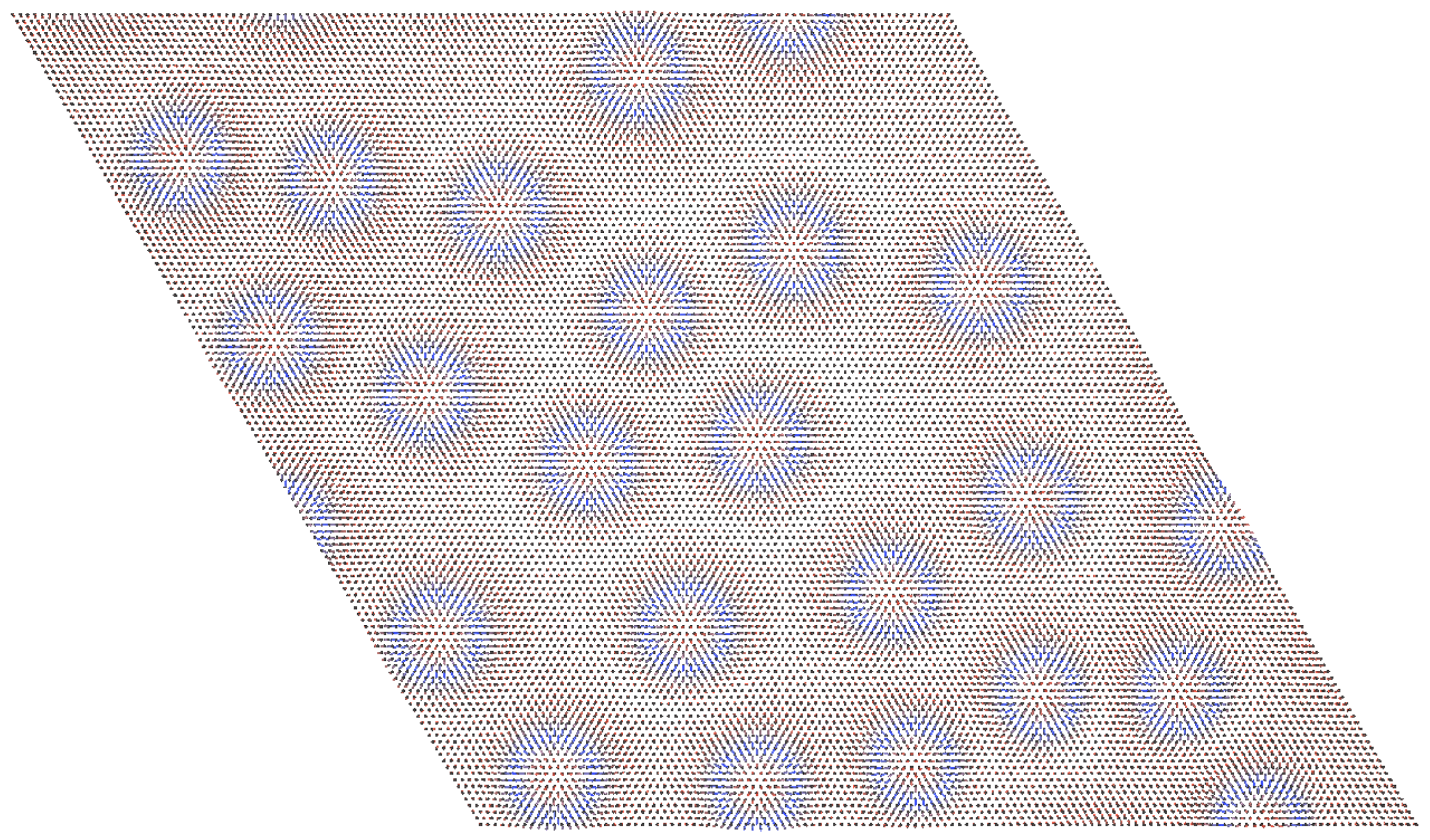}}\\
\subfigure[\label{fig:r10-nobext}] {\includegraphics[scale=0.14]{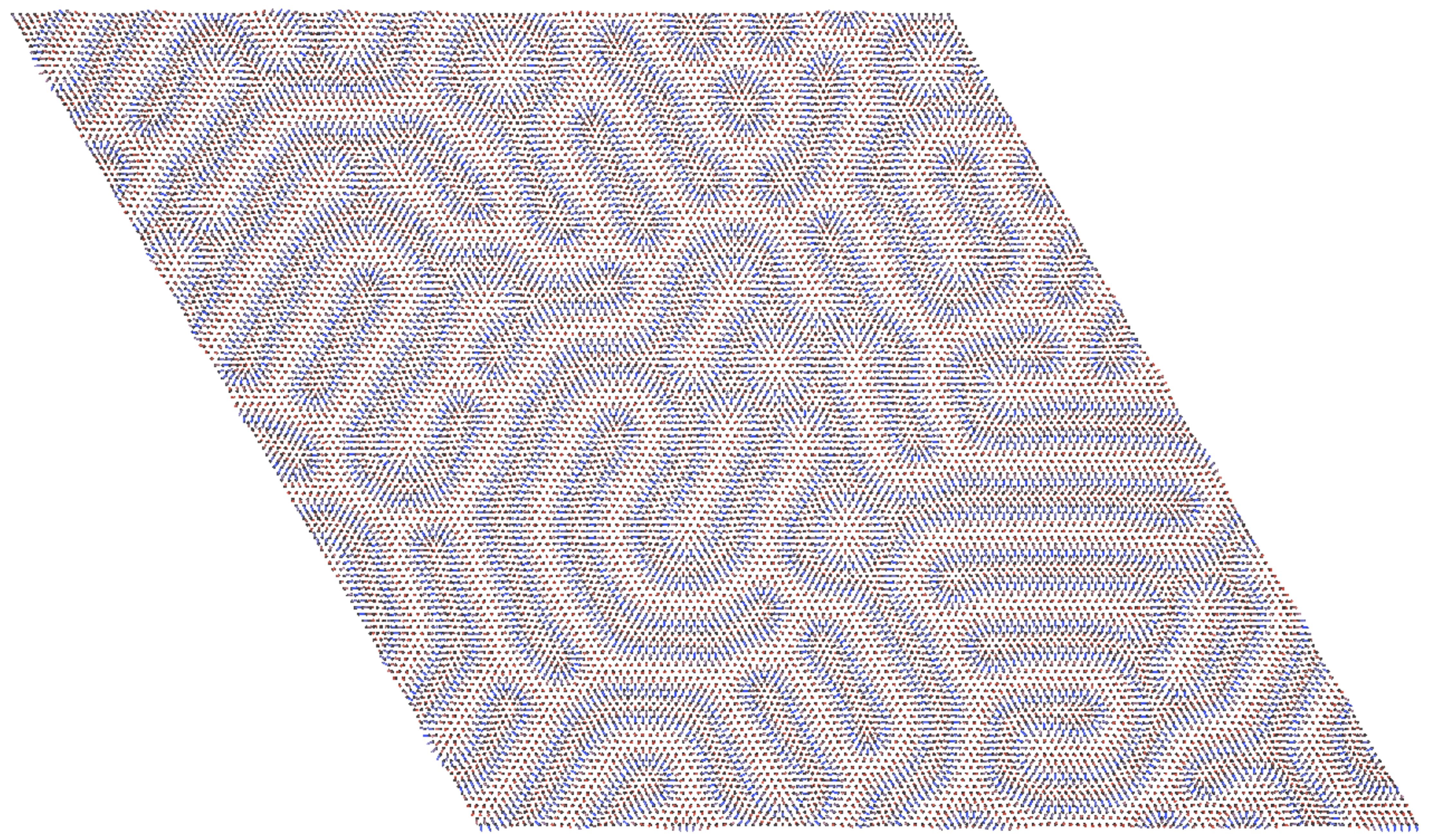}}
\subfigure[\label{fig:r10-bext}] {\includegraphics[scale=0.14]{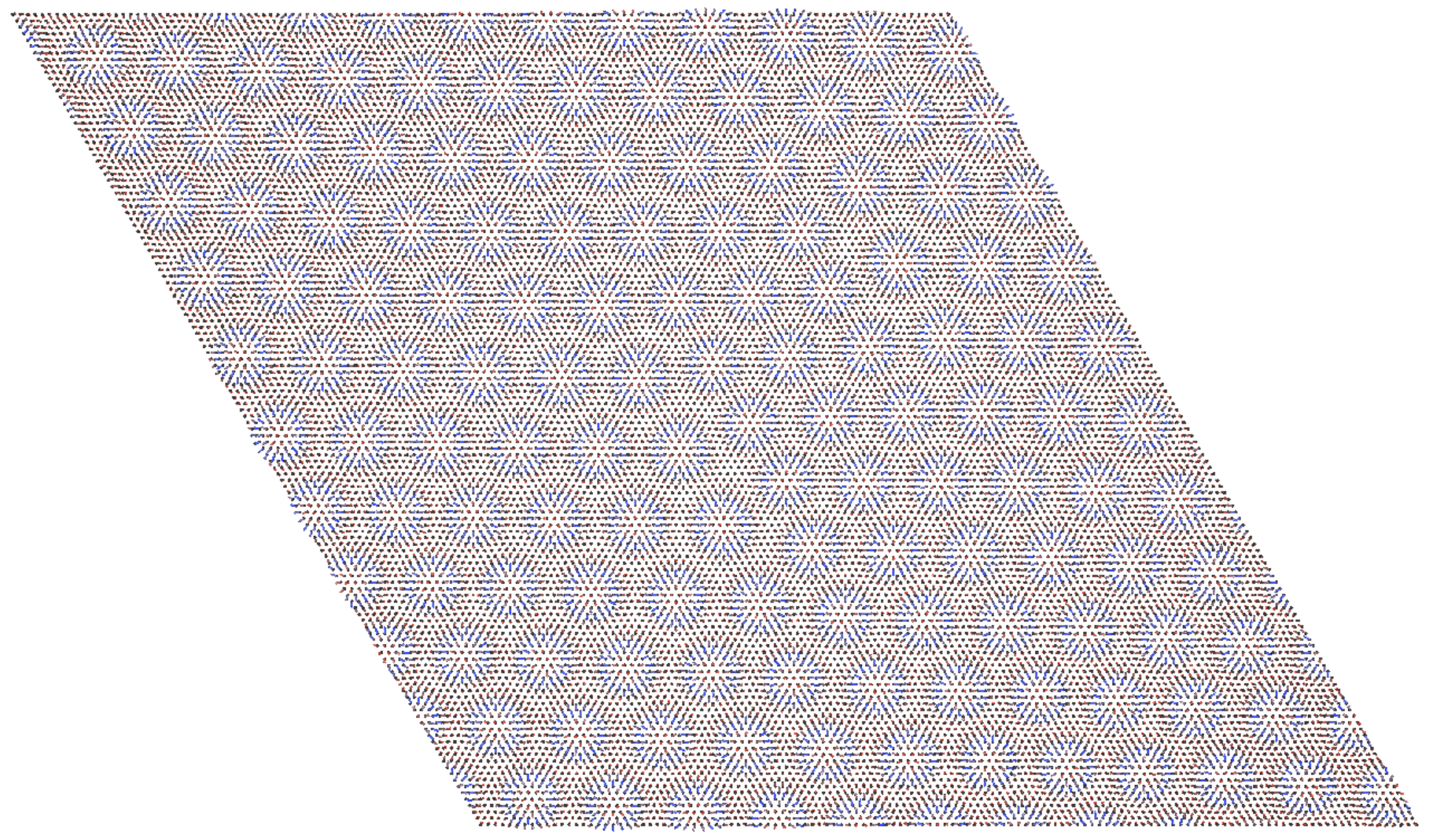}}
\caption{(Color online) ~\subref{fig:r5-nobext} Zero-field ground state spin-spiral configuration of PdFe/Ir(111) system in case of $-5$\% relaxation and ~\subref{fig:r10-nobext} in case of $-10$\% relaxation as obtained from spin-dynamics simulation. External magnetic field leads to skyrmion lattice formation, ~\subref{fig:r5-bext} represents the skyrmion lattice for $-5$\% Fe layer relaxation and ~\subref{fig:r10-bext} for $-10$\% relaxation. Small red and blue arrows indicate magnetic moments with dominating out-of-plane and in-plane components, respectively.}
\label{spindin}
\end{figure*}

Inferring Fig.~\ref{sknum}, in case of $-5$\% layer relaxation the skyrmion lattice phase is formed in the vicinity  of $B_{\rm SkX} \simeq 5$~T. As compared to Fig. \ref{energy}, this is about 3~T higher than that obtained from the energy of frozen spin-configurations in external field.  This quantitative difference can be understood as follows. The spin-configuration of the SS+Sk phase sensitively changes with the change of $B_{ext}$. Clearly, more and more skyrmions are created that considerably reduce the energy of this phase. Therefore, the notion of constant energy for the skyrmion phase in Fig.~\ref{energy} is inconsistent with the results of the spin-dynamics simulations. Similar reasoning applies to explain the high value of the lower border of the FM state (15 T), since the FM state appears via a gradual decrease of $N_{sk}$ with increasing $B_{ext}$. 

From  Fig.~\ref{sknum} we can observe two main features that are remarkably different for the two considered Fe layer relaxations. One of them is that $N_{sk}$ reaches a maximum of 20 for $-5$\% relaxation, while nearly 130 for $-10$\% relaxation. The other one is the much broader range of $B_{ext}$ for $-10$\% relaxation with a
corresponding value of $B_{\rm SkX}=18$~T as opposed to $B_{\rm SkX}=5$~T for $-5$\%.
As for each layer relaxations we used the same lattice (surface area) in our spin-dynamics simulations, it is straightforward to conclude that both features are related to the fact that the size of the individual skyrmions decreases with increasing inward layer relaxation. 

\begin{table}[ht!]
\begin{center}
\begin{tabular}{c c c c c c }
\hline
Relaxation && $D/J$ & $d_{Sk}$(nm) & $d_{Sk-Sk}$(nm) & $B_{\rm SkX}$(T)\\
\hline
\hline
$-5$\%  &&    0.05  &   5.0 &  6.3 &   5 \\
$-6$\%  &&    0.06  &   4.3 &  4.4 &   7 \\
$-7$\%  &&    0.07  &   4.3 &  4.3 &   9 \\
$-8$\%  &&    0.09  &   3.3 &  3.5 &  10 \\
$-9$\%  &&    0.10  &   2.8 &  3.4 &  13 \\
$-10$\% &&    0.19  &   2.6 & 2.3 &  18
\end{tabular}
\caption{Relaxation of the Fe layer, ratio of the magnitudes of the Fe-Fe nearest neighbor DM vector and the nearest neighbor isotropic exchange coupling $(D/J)$, the diameter of skyrmions $(d_{Sk})$, the smallest inter-skyrmion distance ($d_{Sk-Sk}$), and the external magnetic field ($B_{\rm SkX}$) where the skyrmion number takes its maximum. Note that $B_{\rm SkX}$ is determined with an error of 0.5 T.}
\label{diameter}
\end{center}
\end{table}

From the spatial dependence of the normal-to-plane component of the normalized magnetic moments ($s^z=\cos\theta$), we used a domain wall like fit \cite{butenko, dupe, kiselev} to determine the diameter of the skyrmions. At the center of an individual skyrmion $s^z=-1$ ($\theta=\pi$), and $s^z$ approaches $1$ ($\theta=0$) sufficiently far from the skyrmion. In Table \ref{diameter}, for all considered layer relaxations we summarized the determined skyrmion diameters, $d_{Sk}$, the smallest inter-skyrmion distances, $d_{Sk-Sk}$, and the external magnetic fields, $B_{\rm SkX}$, where the skyrmion number is the largest. Note that $d_{Sk-Sk}$ is defined as the distance between the center of the skyrmions. As can be seen, the skyrmion diameter is decreasing with increased inward layer relaxation and correspondingly increased $D/J$, similarly as found for the wave length of the spin-spiral, see Table \ref{tab:wave}. Interestingly, for larger relaxations the spin-spiral wave length and the skyrmion diameter take very similar values. Moreover, we find that for increasing relaxations the packing of  the skyrmion lattice increases: in case of  $-5\%$ relaxation the smallest inter-skyrmion distance is considerably larger than the diameter of the skyrmions, while for $-10\%$ relaxation $d_{Sk} > d_{Sk-Sk}$.
It is also clearly seen in Table \ref{diameter} that $B_{\rm SkX}$ is increasing with increasing inward layer relaxation, i.e. with decreasing skyrmion size.

Finally, in Fig. \ref{spindin} the simulated spin-configurations for the spin-spiral ground states ($B_{ext}=0$) and the skyrmion lattice at $B_{\rm SkX}$ are presented for $-5$\% and $-10$\% Fe layer relaxations.  Noteworthy, the SS ground states are characterized by a strong domain structure which is the consequence of the three-fold degeneracy of the $\vec{q}$ wave vectors of the ground state spin-spirals. It can be noticed that in case of larger size of the skyrmions ($-5$\% relaxation), the high-density skyrmion phase is rather loosely ordered and for small skyrmions ($-10$\% relaxation) a well--ordered skyrmion lattice develops, see also the $d_{Sk-Sk}$ values in Table \ref{diameter}. The origin of the less ordered Sk phase can be purely numerical, as the diameter of the skyrmions is not compatible with the size of the lattice used for the simulation. It is, however, not excluded that this difference can partly be related to different skyrmion-skyrmion interactions, posing a challenging topic for future research. Moreover, one can observe that the area of ferromagnetically ordered spins is much larger for larger sizes of the skyrmions, being a simple space-filling effect. This makes the more relaxed ($-10$\%) film more rigid against applied external fields which explains the larger range of $B_{ext}$ seen in Fig.~\ref{sknum}.

\section{Conclusions}
We investigated the magnetic ground state of PdFe/Ir(111) at different inward layer relaxations and the evolution of the magnetic ground state in external magnetic field. We employed the SCE-RDLM method to obtain spin-model parameters and performed spin-dynamic simulations.
We found that the magnetic ground state without external magnetic field is a spin-spiral in all considered inward layer relaxations. The wave length of the spin-spiral is decreasing with increasing inward layer relaxation due to the increasing ratio of the nearest neighbor DM vector and the isotropic exchange coupling, $D/J$. Applying external magnetic field, skyrmions are created in the system. Based on energy calculations of spin configurations in external magnetic field we obtained good agreement for the phase boundaries with available experiments. Numerically evaluating the skyrmion numbers in spin-dynamics simulations we identified different magnetic phases depending on the magnetic field. We found that the skyrmion diameter and the smallest inter-skyrmion distances decrease with increasing inward layer relaxation and larger external fields are needed to stabilize such skyrmion lattices. Therefore, we conclude that the size of skyrmions and the stabilizing external field can be tuned by manipulating the geometrical structure of the film, e.g. through applying external mechanical strain or electric field or alloying the substrate.\cite{kudrnovsky,ondracek}

\begin{acknowledgments}
This work was supported by the Hungarian Scientific Research Fund projects K84078 and PD83353, and in part by the European Union under FP7 Contract No. NMP3-SL-2012-281043 FEMTOSPIN. KP acknowledges the Bolyai Research Grant of the Hungarian Academy of Sciences. The work of LS was supported by the European Union, co-financed by the European Social Fund, in the framework of T\'AMOP 4.2.4.A/2-11-1-2012-0001 National Excellence Program.
\end{acknowledgments}

\end{document}